\documentclass[11pt]{article}

\usepackage{graphicx}
\usepackage{amsmath,amssymb,amsfonts}
\usepackage{array,tabularx,calc}
\usepackage{multirow}
\usepackage[english]{babel}
\usepackage{natbib}
\usepackage{url}
\usepackage{wrapfig}
\usepackage{float}

\graphicspath{{figs/}}

\newcommand\pic[1]{(Fig.~\ref{#1})}
\newcommand\tab[1]{(Tab.~\ref{#1})}

\newenvironment{conditions}[1][where:]
  {%
   #1\tabularx{\textwidth-\widthof{#1}}[t]{
     >{$}l<{$} @{${}={}$} X@{}
   }%
  }
  {\endtabularx\\[\belowdisplayskip]}

\title{A methodology to rank importance of frequencies and channels in electromyography data with Decision Tree classifiers}

\author{
Albert A. Nasybullin\thanks{a.nasibullin@innopolis.university},\quad
Nursultan Abdullaev\thanks{n.abdullaev@innopolis.university},\quad
Maksim A. Baranov\thanks{m.baranov@innopolis.university},\\
Viacheslav V. Koshman\thanks{v.koshman@innopolis.university},\quad
Vitaly A. Mahonin\thanks{v.mahonin@innopolis.university}\\[6pt]
\textit{Innopolis University}\\
\textit{ul. Universitetskaya 1, Innopolis, 420500 Russia}
}

\date{}

\begin{document}
\maketitle

\begin{abstract}% about 100 words
This study presents a methodology for identifying the most informative frequencies and channels in electromyography (EMG) data to evaluate muscle recovery using Decision Tree classifiers. EMG signals, recorded from the vastus lateralis muscle during squat exercises, were analyzed across varying rest intervals to assess optimal recovery periods. By employing single Decision Tree classifiers, the study enhances interpretability, offering insights into feature importance - essential for applications in medical and sports settings where transparency is critical. The experimental protocol utilized a grid search for hyperparameter tuning and cross-validation to address class imbalance, ultimately achieving a reliable classification of rest intervals based on power spectral density features. The results indicate that a limited subset of highly informative features provides sufficient accuracy, suggesting that streamlined, interpretable models are effective for the evaluation of muscle recovery. This approach can guide future research in developing compact, robust models adapted to EMG-based diagnostics.
\end{abstract}


\section{Introduction}
\setcounter{equation}{0}

Electromyography (EMG) is widely used in biomedical and sports sciences to analyze muscle activity, assess neuromuscular function, and monitor recovery following exertion. EMG signals, reflecting muscle electrical activity, are complex and often noisy, making the identification of informative features essential for effective analysis. Machine learning algorithms, particularly decision trees, have proven effective in EMG classification due to their capacity to capture non-linear relationships and handle complex data.

Recent studies on EMG classification frequently employ tree-based ensemble methods for their robustness and improved accuracy with high-dimensional, noisy data \cite{seyidbayli2020comparison, ramirez2024detection, abdullah2017surface}. However, ensemble models typically lack interpretability --- a critical factor in medical and sports applications where transparency is paramount, as well as for developers seeking insights into model behavior and feature contributions. In this study, we use single Decision Tree classifiers to identify the most significant features in EMG signals. This approach enables interpretable assessment of feature importance, revealing which signal characteristics are most relevant for further analysis and model development.

We aim to identify a set of high-performing features to guide future researchers in designing efficient EMG classification models. Pinpointing key frequencies and channels that impact classification accuracy provides a foundation for subsequent studies to develop compact, effective models tailored to muscle recovery and EMG-based diagnostics.

\section{Data collection and data structure}

The initial sample comprised 14 volunteers, including 11 males and 3 females (78.6\% males and 21.4\% females), aged between 18 and 43 years (mean age = 24.86 years, standard deviation = 7.86). All participants were right-handed, non-smokers, and engaged in regular physical activity. To control for confounding variables, participants were instructed to abstain from alcohol consumption and physical exercise for 48 hours prior to the experiment. They reported being well-rested and fully recovered from their most recent physical activity sessions. Each participant indicated that their caffeine consumption levels were consistent with their typical daily intake. None reported any diagnosed musculoskeletal or central nervous system disorders, nor were they under prescribed medication.

During the exploratory data analysis, three male participants were excluded due to hardware malfunctions and poor data quality. An additional male participant was excluded because of a prior spinal injury. The final sample thus consisted of 10 participants: 7 males and 3 females (70\% males and 30\% females), aged between 18 and 43 years (mean age = 24.5 years, standard deviation = 7.65).

Prior to the commencement of the experiment, all participants were informed about the study's objectives, methodologies, and the potential for minor discomforts.

\subsection{Experimental Design}

\begin{figure}[h]
    \centering
    \includegraphics[width=\textwidth]{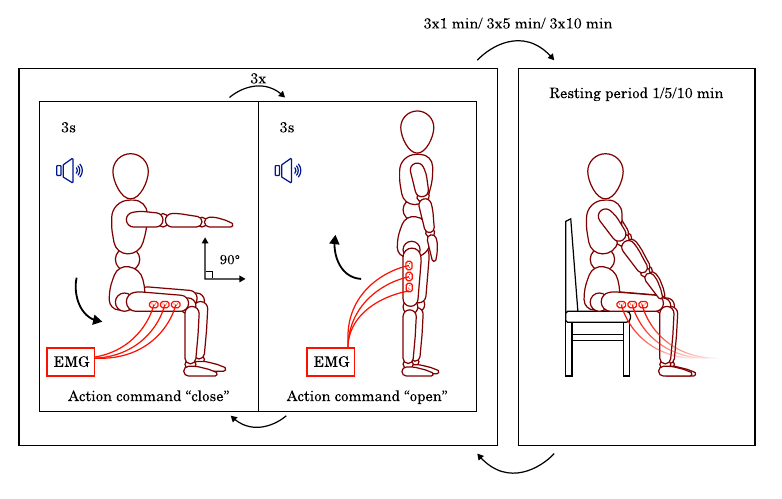}
    \caption{The figure illustrates an experimental protocol for recording muscle activity (EMG) during repeated squat and stand actions. The sequence consists of two main actions, each held for 3 seconds: squatting to a 90$^{\circ}$ knee angle with the command "close" and standing upright with the command "open." This cycle repeats three times within each block. After completing the squat-stand cycle, participants enter a resting period lasting 1, 5, or 10 minutes, depending on the block.}
    \label{fig:exp:process}
\end{figure}

The objective of this experiment is to investigate optimal rest intervals that allow muscle recovery following physical exertion \cite{rest_1_2_3_min, rest_5min_2min}. For the exercise, we selected squats, where participants are required to hold each position (squat or standing) for a duration of three seconds \pic{fig:exp:process}.

\begin{figure}[h]
    \centering
    \includegraphics[width=\textwidth]{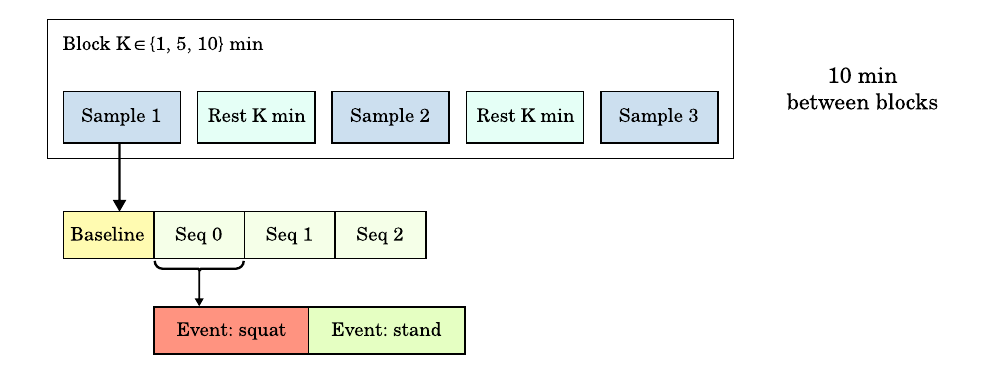}
    \caption{The experimental design consists of three blocks of varying lengths (1, 5, or 10 minutes), each separated by a 10-minute rest interval. Each block includes three sample periods interspersed with rest intervals equal to the block length (K). Following the samples, there is a baseline period, succeeded by three sequential activities (Seq 0, Seq 1, and Seq 2). The sequence includes two specific events: a "squat" event and a "stand" event.}
    \label{fig:exp:pipeline}
\end{figure}

Our experimental setup \pic{fig:exp:pipeline} is divided into three blocks, each exploring different rest intervals: $k \in \{1, 5, 10\}$ minutes \cite{rest_5min_2min}. Between each block, there is a standard 10-minute break, during which the participant rests in a seated position to allow for muscle recovery. Within each block, participants complete three samples, with a rest interval of $k$ minutes between each other.

Each sample consists of a baseline measurement followed by three sequences of actions. In each sequence, there are two main events: a three-second squat hold and a three-second standing hold \pic{fig:exp:process}.

\subsection{Data Acquisition and Preprocessing for EMG Analysis}

\begin{figure}[hbt]
    \centering
    \includegraphics[width=\textwidth]{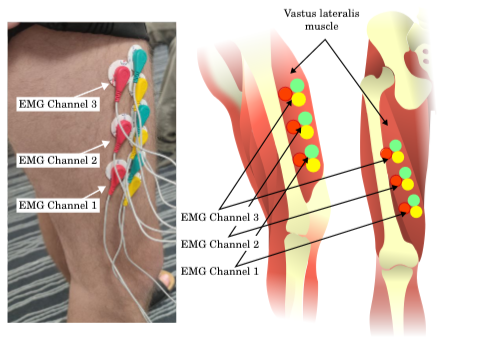}
    \caption{The image shows the electrode placement for EMG recording on the vastus lateralis muscle. Three EMG channels are set up along the thigh, each using a green electrode as the positive (+) terminal, a yellow electrode as the negative (-) terminal, and a red electrode as the reference (ref). The left panel displays the actual electrode placement on a participant's leg, while the right panel illustrates the anatomical positioning on the vastus lateralis muscle, highlighting the alignment of each EMG channel along the muscle.}
    \label{fig:emg}
\end{figure}

Data was recorded using electromyography (EMG) \cite{semg_tutorial_best_practice}. For this purpose, we developed a custom three-channel setup to monitor and analyze the vastus lateralis muscle \pic{fig:emg}.

\subsubsection{System Overview}

The foundation of our system is built on the STM32F401 microcontroller \cite{stm32_ecg}, supported by external electromyography (EMG) signal amplifiers. The microcontroller's integrated analog-to-digital converter (ADC) captures signals, and data transmission occurs at 1 kHz \cite{emg_preprocessing_samplerate}. Direct Memory Access (DMA) is used for efficient data handling, with the data organized into packets and sent via a virtual serial port to the connected computer for further analysis.

\subsubsection{Data Preprocessing}

Upon obtaining the experimental sample data, we perform preprocessing on the EMG signals. The initial step involves baseline correction, where we calculate the average value from a 5-second segment of the recording. Following this, the signal undergoes bandpass filtering within the frequency range of 20 to 500 Hz \cite{semg_why_when_how, semg_acquisition_prosthesis}. Additionally, we employ a notch filter to eliminate power line noise at 50 Hz \cite{semg_acquisition_prosthesis, emg_preprocessing_samplerate}.

\newpage
\subsection{Data processing}

Following the preprocessing of the data, we selected several parameters for the dataset:

\begin{enumerate}
    \item Subject identification number obtained from the questionnaire
    \item Rest period, categorized as 1, 5, or 10 minutes
    \item Experiment number
    \item Sequence, denoted as 0, 1, or 2
    \item Frequency power on channel \(ch\{i\}\_\{freq\}\)Hz, where \(i \in \{1, 2, 3\}\) and \(freq \in \{1, \ldots, 450\}\)
\end{enumerate}

\subsection{Finalized dataset}

The final classification dataset comprises 252 observations, each representing an individual trial from a participant. Each variable corresponds to the power spectral density (PSD) \cite{emg_psd_features, emg_psd_analysis} at specific frequencies ranging from 1 to 450 Hz, recorded from electrodes 1 to 3. Thus, the dataset contains 252 rows and 1 351 columns, with columns representing PSD values across different frequencies and electrodes.

The dataset includes observations from resting intervals of varying durations: 84 observations for 1-minute intervals (34.52\% of the total data), 81 observations for 5-minute intervals (33.33\%), and 87 observations for 10-minute intervals (32.15\%). It is noteworthy that there is a minor imbalance present in the dataset.

% \section{Decision Tree Classifier}

\section{Decision Tree Classifier}

The objective of the machine learning algorithm is to classify data objects according to their respective resting intervals of 1, 5, or 10 minutes. Therefore, the machine learning task is formulated as a classification problem. Decision Tree classifiers were selected due to their high level of interpretability and feature extraction capabilities. Additionally, the small size of the dataset was a significant factor that influenced the choice of model.

While ensemble tree methods are often employed for their robustness and improved accuracy in classification tasks, single Decision Tree classifiers were chosen here due to their simplicity and interpretability, which are valuable in contexts with smaller datasets and a need for transparency in the decision-making process.

Studies such as \cite{seyidbayli2020comparison, ramirez2024detection, abdullah2017surface} highlight the efficacy of ensemble tree classifiers for electromyography (EMG) signal classification, underscoring the potential of tree-based models in similar tasks. Although these works focus on ensemble methods, their results illustrate the foundational strengths of tree-based approaches in EMG signal processing, suggesting that a single Decision Tree classifier can also provide valuable insights, especially when interpretability is prioritized. Thus, the demonstrated success of tree-based models in EMG classification supports our choice to use a Decision Tree classifier in this context.

For this study, we utilized the default implementation of the Decision Tree Classifier from the Scikit-learn machine learning library (version 1.4.2), which is based on the Classification and Regression Trees (CART) algorithm.

\subsection{Grid Search and Cross-Validation}

Given the presence of class imbalance in the dataset, Stratified K-Fold Cross-Validation with three folds was applied. The average F1-weighted score was used to evaluate the classification performance.

Our experimental pipeline consisted of 1 349 individual classification trees. In the initial experiment, all available features (the PSD at specific electrodes) were utilized for classification. Upon completion of the classification, the most important feature was identified based on feature importance evaluated by the mean decrease in impurity. In the subsequent iteration, this top-performing feature was excluded from the training data. Consequently, with each iteration, the number of features used for the next decision tree decreased by one. This procedure continued until only a single feature remained.

For each of the 1 349 decision trees, a grid search was conducted to optimize the hyperparameters. The parameters for the grid search were consistent across all decision trees. The specifications of the decision tree grid search were as follows:

\begin{enumerate}
    \item criterion: \verb|['gini', 'entropy', 'log_loss']|
    \item splitter: \verb|['best', 'random']|
    \item max\_depth: \verb|[None, 10, 20, 30, 50, 100]|
    \item min\_samples\_split: \verb|[2, 5, 10]|
    \item min\_samples\_leaf: \verb|[1, 2, 4]|
    \item max\_features: \verb|[None, 'sqrt', 'log2']|
    \item max\_leaf\_nodes: \verb|[None, 10, 20]|
    \item min\_weight\_fraction\_leaf: \verb|[0.0, 0.1, 0.2]|
\end{enumerate}

\begin{figure}[h]
    \centering
    \includegraphics[width=0.95\textwidth]{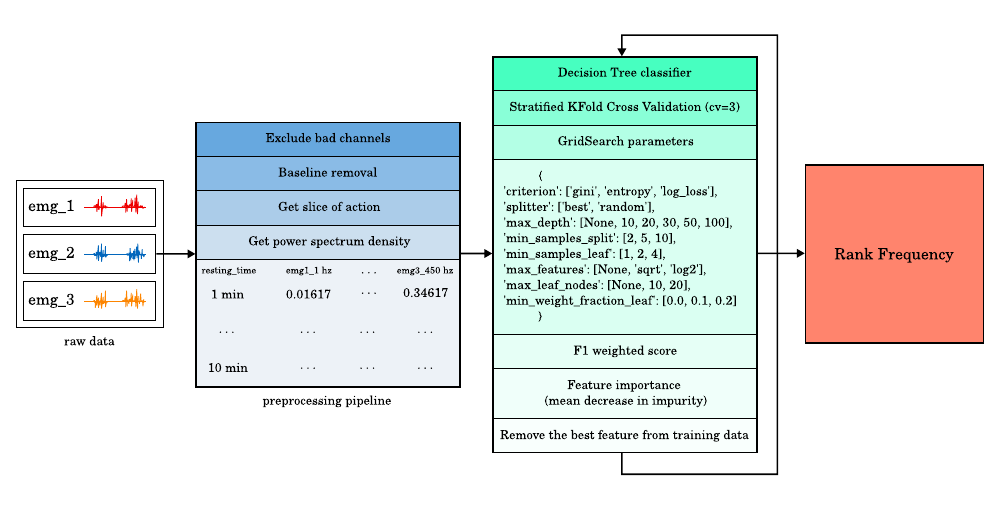}
    \caption{Figure illustrating the analysis pipeline for EMG signal processing and feature selection. Raw EMG data from multiple channels undergoes preprocessing, including bad channel exclusion, baseline removal, and power spectral density calculation. Features are extracted and classified using a Decision Tree classifier with stratified K-fold cross-validation and GridSearch for hyperparameter tuning. Feature importance is assessed, and the least informative features are iteratively removed to rank feature frequency in the final model.}
    \label{fig:ml_pipeline}
\end{figure}

The overall pipeline \pic{fig:ml_pipeline} consists of three main steps: raw data collection, preprocessing, and a classification experiment. The results of the experiment will be utilized in the subsequent chapters, where we introduce the "Rank Frequency" method to evaluate the importance of specific electrodes and frequencies in EMG tasks.

\subsection{Results}

In this study, the highest-performing classifier achieved an average F1-weighted score of 0.4974 \pic{fig:all_classifiers_report}, in comparison to a baseline score of 0.33. Notably, half of the classification models exceeded an average F1-weighted score of 0.42 \pic{fig:box_plot_classification_report}.

\begin{figure}[h]
    \centering
    \begin{minipage}{0.48\textwidth}
        \centering
        \includegraphics[width=\textwidth]{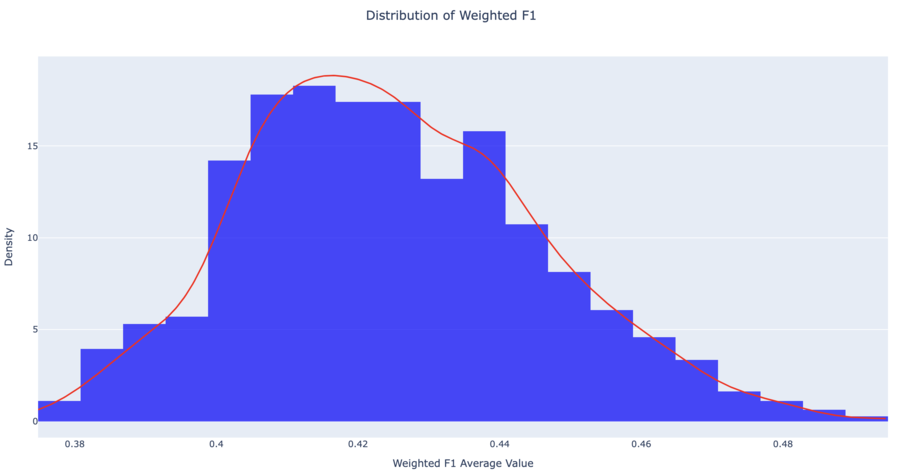}
    \end{minipage}
    \hfill
    \begin{minipage}{0.48\textwidth}
        \centering
        \includegraphics[width=\textwidth]{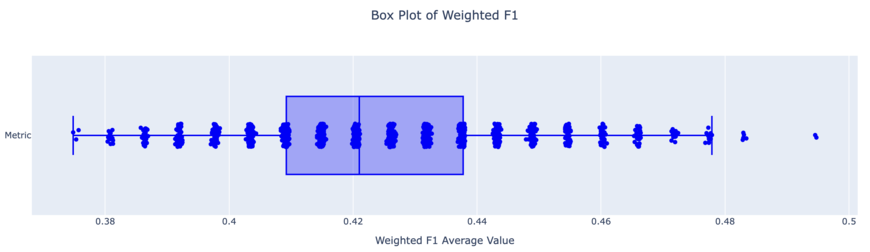}
    \end{minipage}
    \label{fig:box_plot_classification_report}
    \label{fig:all_classifiers_report}

    \caption{The distribution of the weighted F1 average score across all trained decision trees indicates that a subset of classifiers significantly outperforms the rest.}
    
\end{figure}

To investigate the relationship between the number of features (particularly informative features) and the average F1-weighted score, a linear regression model was employed. Here, ``informative features'' are defined as those with non-zero importance values, signifying their contribution to the predictive accuracy of the decision tree. A 95\% confidence interval (CI) was applied around the regression line to account for uncertainty in model predictions. The shaded region illustrates the 95\% CI, indicating a 95\% probability that the true mean response falls within this interval for a given feature count. This method enhances interpretability by highlighting the precision of the regression estimates.

To assess the normality of the data distribution, three statistical tests were conducted: the Kolmogorov-Smirnov, Shapiro-Wilk, and Anderson-Darling tests. Results from all tests suggested significant deviation from normality, leading to the rejection of the null hypothesis that the dataset is normally distributed \tab{tab:normality_test_results}.

\begin{table}[h!]
    \centering
    \resizebox{\textwidth}{!}{\begin{tabular}{|c|c|c|c|}
        \hline
        \textbf{Test} & \textbf{Statistic} & \textbf{P-value / Critical Value} & \textbf{Conclusion} \\
        \hline
        Kolmogorov-Smirnov & 0.0761 & $3.05 \times 10^{-7}$ & Reject the null hypothesis distributed. \\
        \hline
        Shapiro-Wilk & 0.9868 & $1.04 \times 10^{-9}$ & Reject the null hypothesis. \\
        \hline
        \multirow{5}{*}{Anderson-Darling} & \multirow{5}{*}{6.1204} & 15.0\%: 0.574 & Reject the null hypothesis \\
        & & 10.0\%: 0.654 & Reject the null hypothesis \\
        & & 5.0\%: 0.785 & Reject the null hypothesis \\
        & & 2.5\%: 0.915 & Reject the null hypothesis \\
        & & 1.0\%: 1.089 & Reject the null hypothesis \\
        \hline
    \end{tabular}}\caption{Normality tests for the distribution of classification results across decision trees.}
    \label{tab:normality_test_results}
\end{table}

The findings underscore a relationship between the total number of features used by a classifier and the performance quality of a decision tree model \pic{fig:classification_all_features}. However, when focusing exclusively on ``informative features'' with non-zero importance values, this relationship does not hold \pic{fig:classification_important_features}. This indicates that simply increasing the number of such features does not directly correlate with improvements in predictive performance.

\begin{figure}[h]
    \centering
    \includegraphics[width=\textwidth]{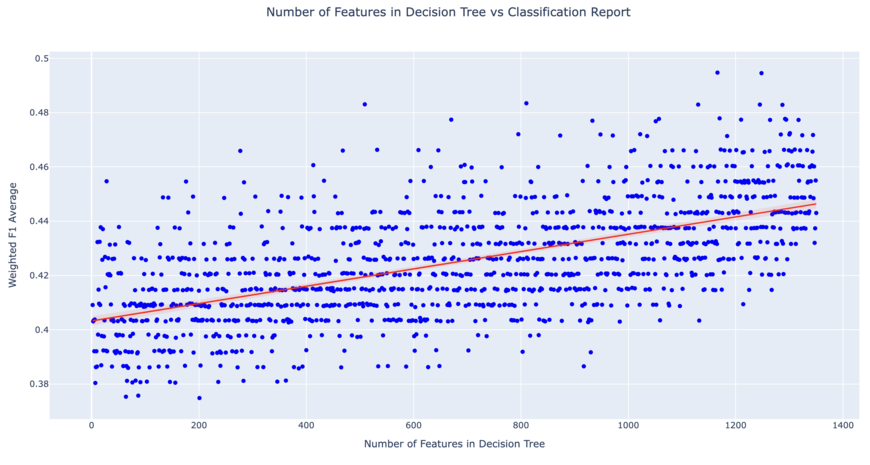}
    \caption{A notable trend indicates that classification quality improves as the number of features increases.}
    \label{fig:classification_all_features}
    
\end{figure}

\begin{figure}[h]
    \centering
    \includegraphics[width=\textwidth]{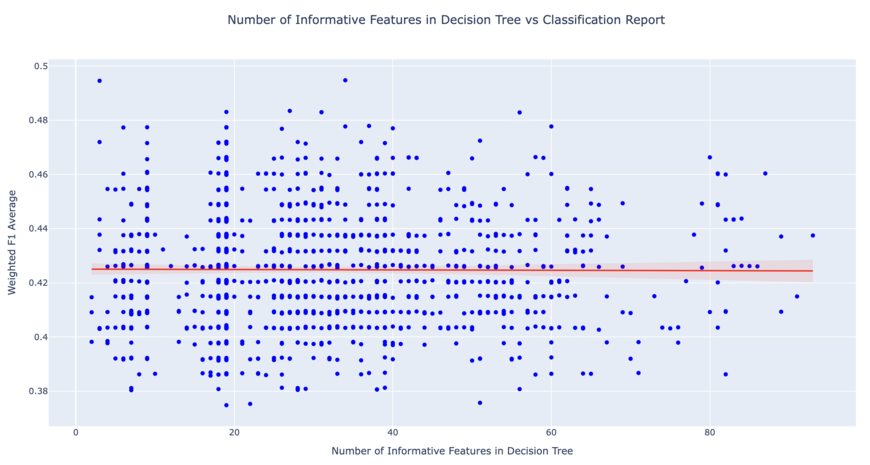}
    \caption{When considering only features with non-zero importance, no dependency is observed between the number of features in the decision tree and the classification quality.}
    \label{fig:classification_important_features}
\end{figure}

\section{EMG Rank Frequency}

Each trained decision tree classifier operates with a distinct subset of features. In this chapter, we propose a method to evaluate the stability of specific electrodes and frequencies in terms of classification robustness and performance.

The initial step involves selecting the top 10\% of classifiers based on the primary classification metric, resulting in a subset of 135 high-performing decision tree models. Drawing inspiration from search ranking algorithms, we introduce feature ranking metrics to assess the importance of each feature. 

The proposed ranking function \eqref{eq1} incorporates several key aspects deemed essential for ensuring stability, classification performance quality, and interpretability of EMG data. Specifically, the ranking function evaluates: the predictive performance of the model, as measured by cross-validation scores; the proportion of informative (non-zero) features contributing to classification; the rank of an individual decision tree relative to other classifiers; and the rank of a particular feature relative to other features within each decision tree.

% \newpage
\begin{equation}
\resizebox{.9\textwidth}{!}{$\displaystyle \mathrm{FeatureRank} = \frac{I(f_i)}{4} \cdot \sum_{i=1}^{N} \left( \frac{\mathrm{mean}(f1(cv)_i)}{\max_{x \in f1(cv)_i} x} + \frac{|f(x \neq 0)_i|}{|f_i|} + \frac{1}{\mathrm{dt\_rank}_i} + \frac{1}{\mathrm{f\_rank}_i}\right)$}
\label{eq1}
\end{equation}

\begin{conditions}
N & total number of decision trees analysed for ranking \\
i & index of a particular decision tree \\
I(f_1) & indicatory function (0 if a feature is not "informative", 1 if a feature is "informative")\\
f1(cv)_i & set of cross-validation scores for a particular decision tree \\
|f(x \neq 0)_i| & total number of features in particular decision tree with non-zero importance \\
|f_i| & total number of features in particular decision tree \\
dt\_rank_i & rank of particular decision tree (defined as a position among all decision trees by target metric) \\
f\_rank_i & rank of a feature in a particular decision tree \\
\end{conditions}

% The Lagrangian function for a harmonic oscillator is given in
% Eq.~\eqref{eq1}.

\begin{figure}[H]
    \centering
    \includegraphics[width=\textwidth]{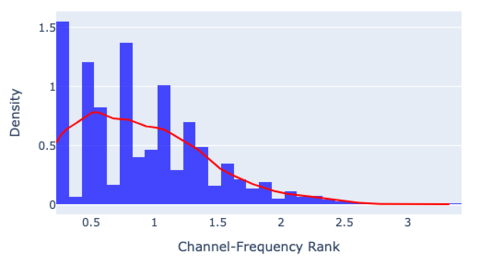}
    \caption{The distribution of Channel-Frequency ranks across all trained decision trees and their respective features.}
    \label{fig:rank_features}
\end{figure}

\begin{figure}[H]
    \centering
    \includegraphics[width=\textwidth]{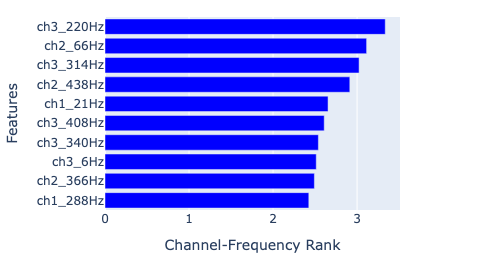}
    \caption{The ten highest-ranked features for muscular recovery period classification encompass both high and low frequency components. Notably, electrode 3 accounts for six of these top-ranked features.}
    \label{fig:10best}
\end{figure}

The Rank Frequency method was applied to all trained decision trees in this study. Of the 1,350 features analyzed, 97 received a rank of zero, 827 were ranked below one, and only three features achieved a rank exceeding three \pic{fig:rank_features}.

The Rank Frequency method was applied to all trained decision trees in this study. Among the 1,350 features analyzed, 97 received a rank of zero, 827 were ranked below one, and only three features achieved a rank exceeding three. We propose the Rank Frequency method as an analytical and discovery tool for researchers and EMG enthusiasts, as it effectively distinguishes a subset of EMG features with considerable variability \pic{fig:rank_features}. Furthermore, we identify and present the top ten highest-performing features for classification tasks related to muscle recovery period assessment \pic{fig:10best}.

\section{Conclusion}

This study illustrates the application of decision tree classifiers as the foundation of a newly proposed method for ranking the importance of frequencies and channels in EMG data to assess muscle recovery. By analyzing the power spectral density of EMG signals across various resting intervals, this method identifies key features that significantly enhance classification quality, providing a transparent approach to feature selection. The proposed method demonstrates a high degree of interpretability, which is particularly valuable in clinical and sports contexts. Findings indicate that a limited number of highly informative features are sufficient for effective classification, reducing the complexity of models while maintaining accuracy.

We anticipate that the proposed method will reduce data redundancy and enhance research opportunities for future investigations in fields such as medical and sports studies, the design of weak EMG classifiers, and the development of specialized high-precision hardware systems, including field-programmable gate arrays (FPGAs).

Future research could expand on these findings by exploring ensemble methods or incorporating additional preprocessing techniques to further enhance model robustness across a wider range of applications. The primary focus of the current study was on experimental, hardware, and software design. We acknowledge that the number of participants in this study may be limited compared to other studies. As a potential direction for future work, the evaluation and refinement of the proposed method on a larger-scale open-source EMG dataset is planned, with the aim of comparing ranked frequencies and channels to results obtained in previous research.

To support openness and reproducibility in scientific research, we share all source data and code generated during the study, with the exception of personal and sensitive data \footnote{\url{https://github.com/Slauva/Ranking-frequencies-and-channels-of-EMG}}.

\bigskip
\section*{Acknowledgements}

We would like to express our gratitude to all participants who contributed to the experiments. Additionally, we extend our thanks to the artist Julia Ivshina for her invaluable work in creating the diagrams and illustrations for this article.

\end{document}